# EVOLUTIONARY CENTRALITY AND MAXIMAL CLIQUES IN MOBILE SOCIAL NETWORKS

Heba Elgazzar[1] and Adel Elmaghraby[2]

School of Engineering & Information Systems, Morehead State University
Morehead, KY, USA
h.elgazzar@moreheadstate.edu

[2] Computer Engineering and Computer Science Department, University of Louisville
Louisville, KY, USA
adel.elmaghraby@louisville.edu

## ABSTRACT

*This paper introduces an evolutionary approach to enhance the process of finding central nodes in mobile networks. This can provide essential information and important applications in mobile and social networks. This evolutionary approach considers the dynamics of the network and takes into consideration the central nodes from previous time slots. We also study the applicability of maximal cliques algorithms in mobile social networks and how it can be used to find the central nodes based on the discovered maximal cliques. The experimental results are promising and show a significant enhancement in finding the central nodes.*

## 1. INTRODUCTION

Centrality in social networks is an important measure of the influence of the node in the network [1-6]. Finding these central nodes that have important effects on other nodes will have many important applications in network science. One application in healthcare is discovering these central nodes that transmit disease among the nodes that represent connected communities. Finding clusters that present communities in social networks is also one of the important problems [7-11].

In this paper, we focus on the design and analysis of evolutionary centrality algorithms that can be used effectively to analyze the dynamics of mobile networks to find central nodes. There is a need to understand the behavior of mobile networks and to find central nodes dynamically. This can enhance the provided services and introduce new applications that meet the need of communities in mobile networks. The goal of this research is to develop algorithms that can be used for evolutionary centrality in mobile social networks by utilizing mobility data that represents the dynamic behavior of the mobile networks.

The remainder of this paper is organized as follows. Section 2 gives an overview of network representation. Section 3 discusses different centrality measures in social networks and the proposed evolutionary approach. Section 4 discusses one of the maximal cliques algorithm that can be applied to mobile social networks. The MIT mobility dataset that was used in our experiments are discussed in section 5. Sections 6 and 7 discuss the results of several experiments that we conducted to evaluate the evolutionary centrality and to discover maximal cliques. Finally, conclusion and future work are given at the end of this paper.

## 2. NETWORK REPRESENTATION

The work in this paper is related to network science, which is one of the important fields in computer science. Network science focuses on the study of the theoretical foundations of network structure, network dynamics, network behavior, and the application of network science to different types of networks [12]. Network science techniques can be applied to different types of networks including mobile social networks, biological networks, power grids and other types of networks. We can represent any network in terms of nodes, edges between the nodes, and algorithms that can be applied to the nodes. Network nodes represent the main entities in the network that interact with each other. An edge or a link represents an interaction or a relationship between two nodes. We can define any network as shown below [12]:

$$G(t) = \{ N(t); L(t); f(t) : J(t) \} \qquad (1)$$

where:
  *N(t): is nodes or vertices at time t*
  *L(t): is links or edges at time t*
  *f(t): N X N mapping function that connects node pairs*
  *J(t): Algorithm for describing behaviors of nodes and links versus time.*

The above definition gives an important representation for the topology of the networks which can be changed dynamically with time and connects the different entities in the networks. Algorithms can be used to understand the behavior and the dynamics of the networks. In this paper, we will focus on mobile nodes in mobile networks where the network nodes are social communicating and the topology is changing with time because the nodes are not stationary and the communication links are dynamically changing between nodes. An example of this mobile network is shown in Figure 1.

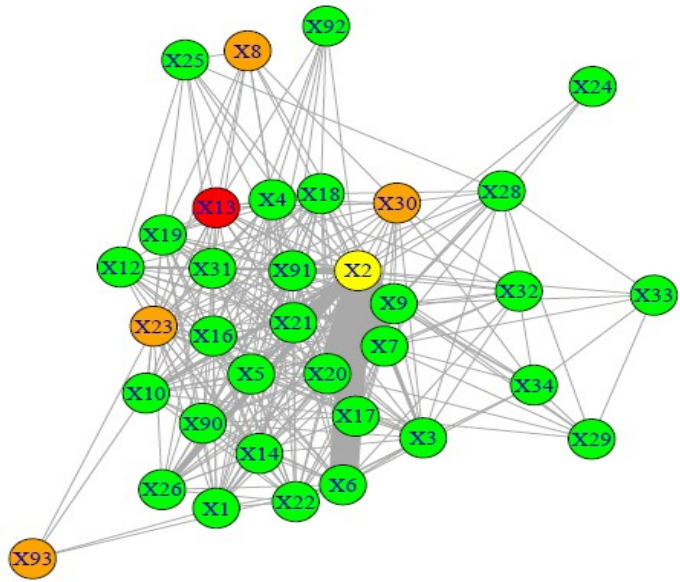

Figure 1.  Example of network representation

## 3. CENTRALITY IN SOCIAL NETWORKS

Centrality in social networks is a measure of the importance of the node in the network. This is a very useful measure and has important applications. This can be used to discover central nodes that have important effect on other nodes.

There are several techniques that can be used to measure the centrality in social networks. In this paper, we focus on the following similarity measures that can be used effectively for mobile networks where the network is changing dynamically over time [1-6]:

### 3.1. Degree Centrality

This is a simple measure and it is based on counting the number of neighbor nodes. This degree centrality, CD, of a node N can be expressed as the degree of the vertex N in the connected graph that represents the network:

$$CD(N) = deg(N) \qquad (2)$$

where *deg(N)* returns the number of neighbor nodes that are connected to the node *N*.

### 3.2. Closeness Centrality

This measures the centrality based on the distance from the node to the neighbor nodes. The sum of the shortest distance to all other nodes is used. This measure is useful to show how fast it will take to reach other nodes. This is very useful in broadcasting and healthcare applications.

$$C_c(N) = \frac{1}{\sum_{v \neq N} d(N,V)} \qquad (3)$$

where d(N,V) *is the shortest distance between nodes* N *and* V

### 3.3. Betweenness Centrality

This measures the centrality based on the number of times the node acts as a bridge between other two nodes based on the distances between these nodes. The betweenness centrality for a node N can be calculated according to the following steps:

1. Start with $C_B(N) = 0$
2. For each two other nodes ($N_1$ and $N_2$) find the shortest path between them:
    a. If this shortest path passes through N, find p, the percentage of that path that passes through N.
    b. $C_B(N) = C_B(N) + p$
3. Repeat step 2 for all pairs of nodes.

### 3.4. Eigenvector Centrality

This measure is based on the idea of considering a node as an important node if it is connected to other important nodes. It is based on counting the number of neighbor nodes. The vector of centralities, C, can be found based on the adjacency matrix *A* and a constant $\lambda$ as follows:

$$\lambda . C = A . C \qquad (4)$$

where *A* is an $n \times n$ adjacency matrix and *n* is number of all nodes in the network. Any element $a_{ij}$ in *A* represents the connection (edge) between the nodes *i* and *j*. Element $a_{ij}$ will be zero if there is no edge between nodes *i* and *j*. *C* is the eigenvector of *A* with an eigen value of $\lambda$.

### 3.5. PageRank Centrality

This centrality measure algorithm was developed by Google to determine the importance of webpages based on counting the number of links to the page and the quality of these links [6]. The PageRank, *PR(N)*, for any node, *N*, in this algorithm is based on PageRank *PR(J)* for every other node J connected to *N* and the number of links (edges) of *J*. This can be expressed in a simple form as:

$$PR(N) = \sum_{for\ all\ nodes\ J\ conncetd\ to\ N} \frac{PR(J)}{L(J)} \qquad (5)$$

*where:*
  *PR(N) is the PageRank of the node N*
  *PR(J) is PageRank of the node J that is connected to N*
  *L(J) is the number of links of the node J*

In this case, the PageRank measure for each node will be initialized to 1/n where n is the total number of nodes in the network.

### 3.6. Proposed Evolutionary Centrality

We introduce an evolutionary approach to enhance the process of finding central nodes. The evolutionary centrality will be calculated based on the connections between the nodes in the network at the current time slot, t, and at the previous time slot, t-1. This evolutionary approach considers the dynamics of the network and takes into consideration the central nodes from previous time slots. The factor α will be used to calculate the updated adjacency matrix between the nodes (which represents the connections between edges) by considering the connections from current and previous time slot:

$$A'_t = (1 - \alpha)A_t + \alpha A_{t-1} \qquad (6)$$

*where:*
  *$A_t$ is the adjacency matrix at time t*
  *$A'_t$ is the updated adjacency matrix after temporal smoothing*

This approach is applicable to all the above centrality measures considering the dynamics of the network based on the updated adjacency matrix that considers the dynamic network structure at different time slots.

## 4. Maximal Cliques

A clique in social networks represents a group of people or nodes that are connected in a complete graph where every node is connected to each other. An example of a clique is shown in Fig. 2. A k-clique is a clique of order k where it has k nodes. The clique shown in Fig. 2 is an example of a 13- clique. A maximal clique is a clique that is not a subset of any other clique. The graph shown in Figure 2 is a maximal clique while any subgraph of this graph is a clique.

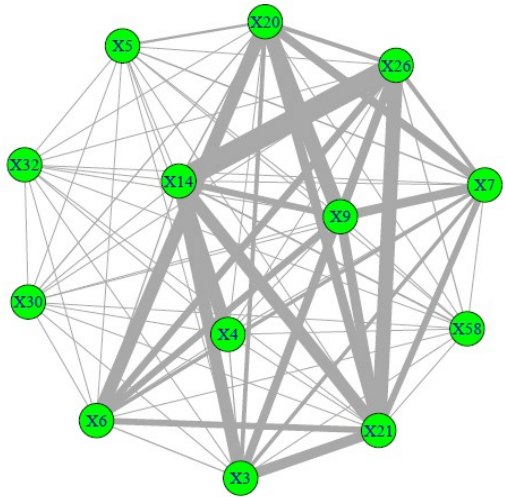

Figure 2. Example of a clique

The well-known Bron-Kerbosch algorithm [11] was used in this problem in a recursive manner to find the maximal cliques for mobile social networks. The general algorithm requires an undirected graph with no self-edges which is applicable in mobile social networks where the link between two nodes is undirected and there is no edges from one node to itself [11]. Suppose that there are three sets R, P, and X where [11]:

*R: represents the set of vertices that represent a maximal clique or can be extended to a maximal clique.*

*P: represents the set of vertices that are connected to all vertices in R and can be added to R to make a larger clique.*

*X: represents the set of vertices connected to all vertices in R but excluded from being added to R.*

*N(v): returns a list of vertices that are neighbors to v.*

The Bron-Kerbosch algorithm starts by initializing R and X to empty sets and P to the set of the vertices in the graph. The algorithm makes recursive calls to consider every vertex in P in turn. The general algorithm is as follows [11]:

```
BronKerbosch(R, P, X):
   1. if P and X are empty:
            Report R as a maximal clique
   2. for each vertex v in P:
BronKerbosch(R∪{v},P∩N(v),X∩N(v))
         P = P\{v}
         X = X∪{v}
```

This algorithm is one of the most effective algorithms to find clique communities, and therefore, it was selected to find the maximal cliques in our problem.

## 5. MIT REALITY MINING DATASET

MIT reality mining dataset [13] was used in our experiments in the next sections to test the proposed evolutionary centrality and to find the maximal cliques. This dataset has records for 106 users and it was recorded for a period of 9 months. The dataset was available as MATLAB data. This data has a variety of information including mobility data survey data collected for mobile users. MATLAB toolboxes were used to help in implementing the evolutionary centrality.

We considered the mobility features in the datasets to find friendship among users. This data includes MAC address data, location data, and time stamps. This data is recorded for all mobile users. Our analysis in the next sections is based on the fact that the MAC address can be discovered by a periodic Bluetooth scan performed by another phone, and this is indicative of the fact that the two users' phones are within 5-10 meters of each other. A similarity matrix was used to record the similarity between users which represents the number of intervals where they were in physical proximity for different weeks. This dataset was used in the next sections to find central nodes using the proposed evolutionary centrality and to find the maximal cliques.

## 6. EXPERIMENTAL RESULTS FOR CENTRALITY

Several experiments have been conducted to find the centrality in the mobile social networks using different centrality measures. In this paper, we used the results of the PageRank centrality and eigenvector centrality to show the effectiveness of using the evolutionary approach.

Figure 3 shows an example of the results before and after applying the evolutionary centrality using the eigenvector centrality measure for time slot 46. Figure 3 (a) shows the eigenvector centrality at time slot 45 and Figure 3 (b) shows the eigenvector centrality at time slot 46 before applying the evolutionary centrality. Fig. 3(c) shows the results of applying the evolutionary centrality at time slot 46 by considering the network connections at time slot 45 and time slot 46. The results of the centrality scores in Figure 3 (c) show a significant enhancement in finding the central nodes using the evolutionary centrality. For example, Figure 3 (b and c) shows that the relative centrality score for nodes 17 was increased significantly at the current time slot (t=46) by considering the network connections at the previous time slot (t=45). Figure3 (a) shows the centrality for node 17, and it was the main central point at that time slot. Also, the relative centrality score for node 44 was increased significantly, because it continues to represent a high level of centrality over time as indicated in Figure 3.

Additional experiments were conducted to apply the evolutionary centrality for the MIT mobility dataset. The results of the scores for different time slots for PageRank centrality before applying the evolutionary approach are shown in Figure 4. The results of evolutionary centrality using the PageRank Centrality measure are shown in Figure 5 for different time slots. These results are shown for α factor of 0.5. The results show a significant improvement in finding central nodes by applying the evolutionary centrality approach. The nodes centrality measures were adjusted to reflect the centrality over time for different nodes. This is very important in many applications where the network is changing dynamically.

Several experiments were also conducted for all other centrality measures (degree centrality, closeness centrality, and betweenness centrality). The results of degree centrality show that central nodes can be detected in the network directly, and a threshold can be used to determine the effective central nodes. The results of closeness centrality show that central nodes that can be used to reach other nodes in a fast way based on the distances. The results of the scores for the betweenness centrality show that central nodes that can be used effectively to connect other pairs of nodes. In this paper, we focused on the results of the evolutionary PageRank centrality, which is very applicable to dynamic networks, because it considers the updated links.

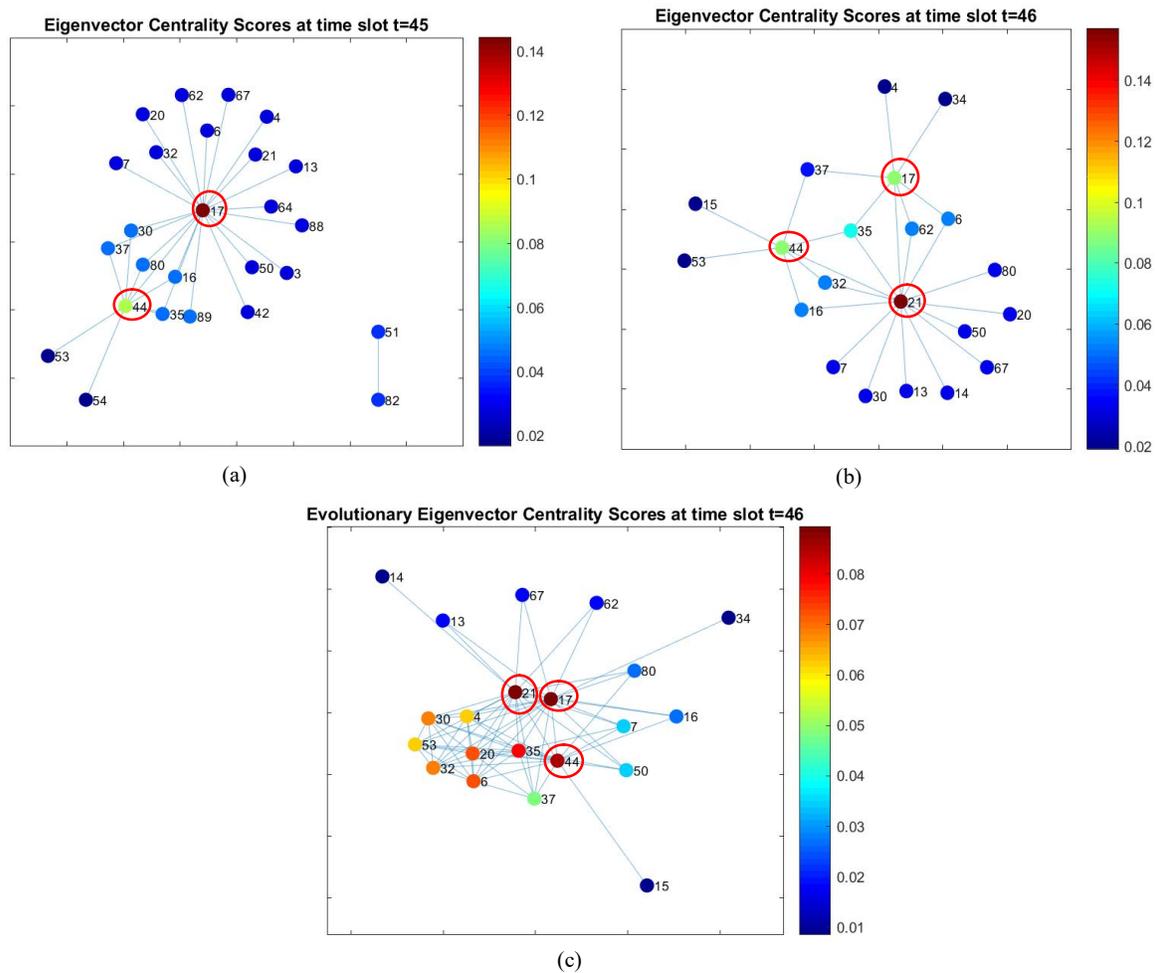

Figure 3. The results applying the evolutionary centrality for Eigenvector Centrality at time slot 46 with α = 0.5 of

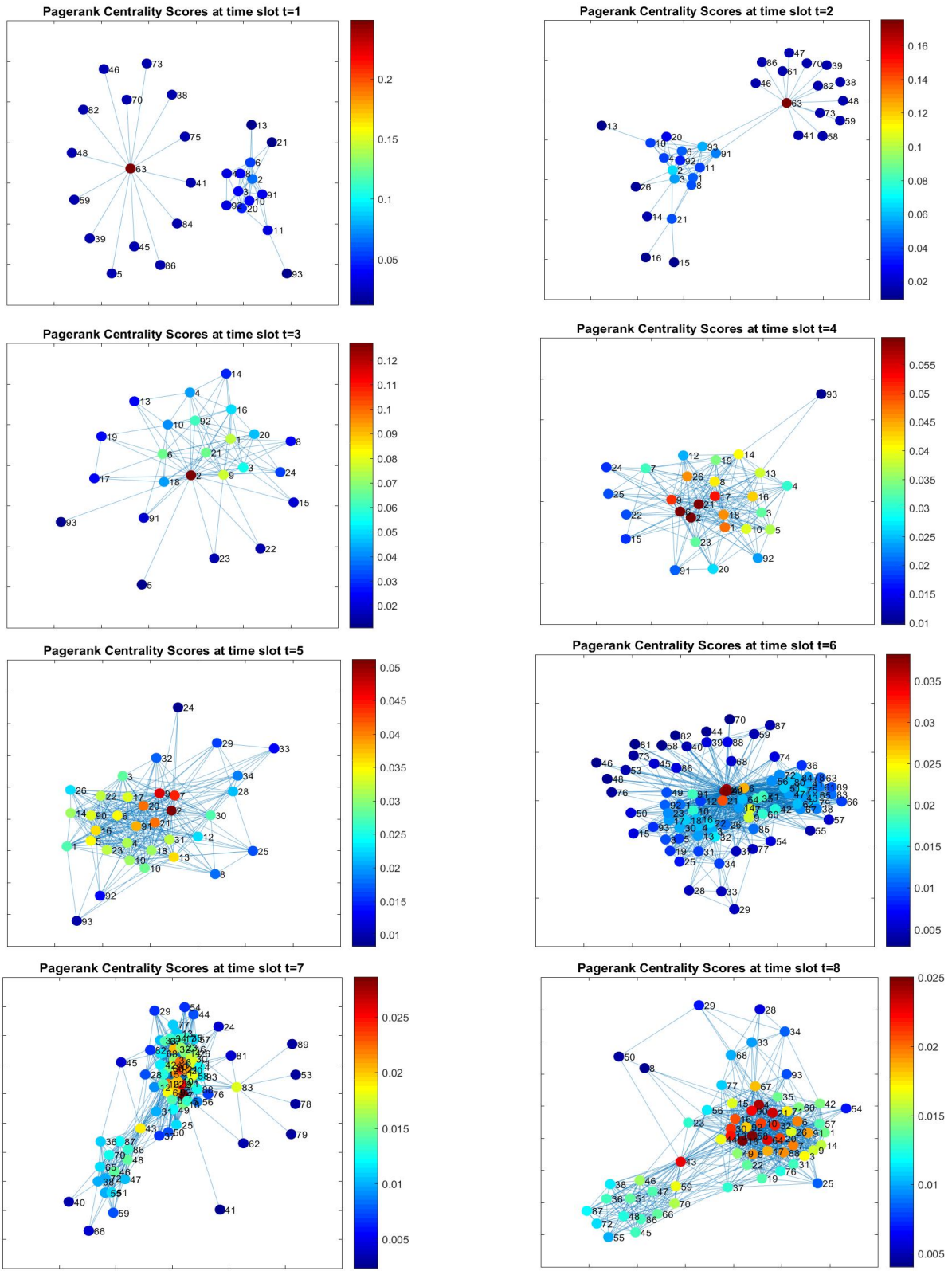

Figure 4. PageRank Centrality Scores for different time slots

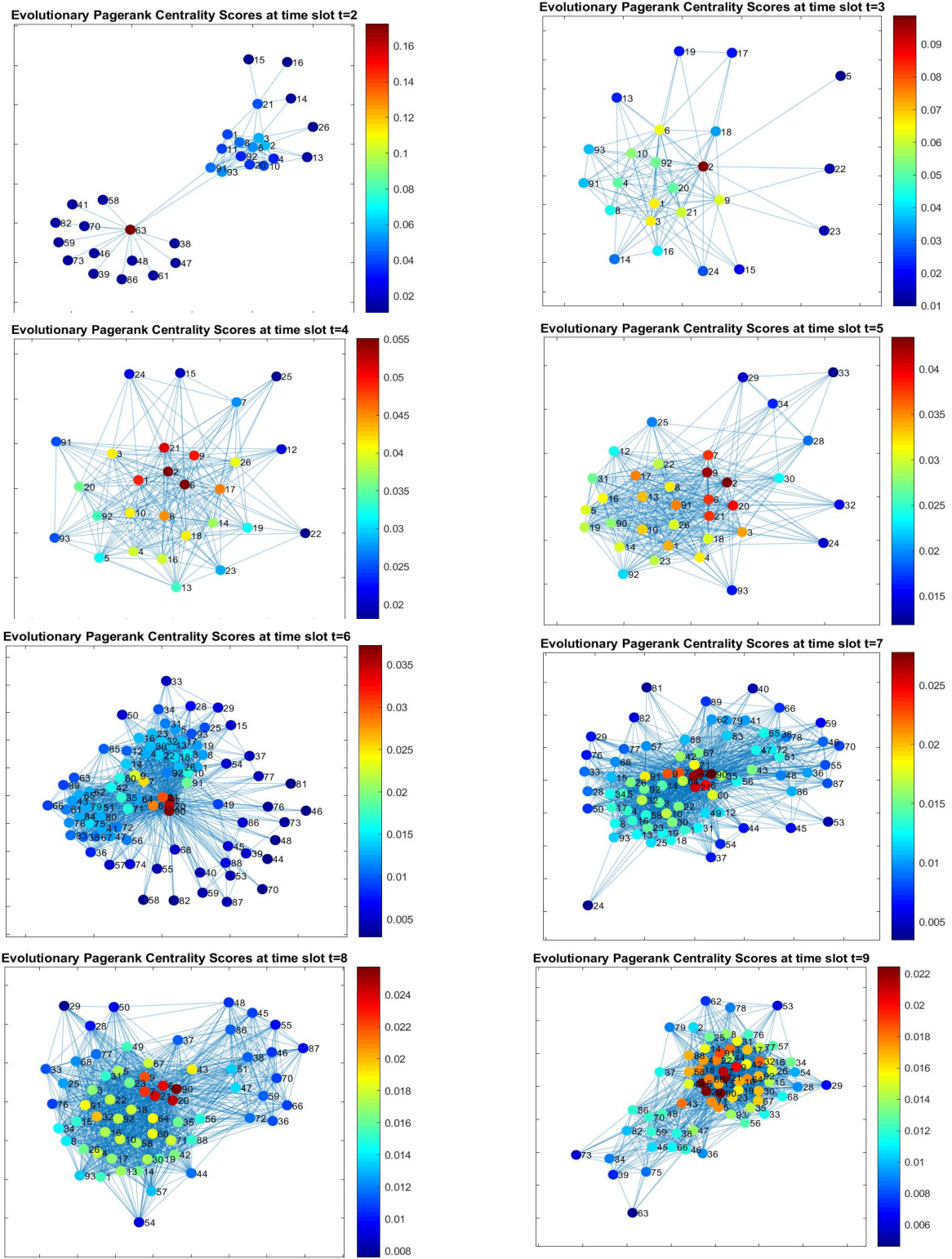

Figure 5. Evolutionary PageRank Centrality Scores for different time slots with α = 0.5

## 7. EXPERIMENTAL RESULTS FOR MAXIMAL CLIQUES

We applied the Bron-Kerbosch algorithm for the MIT reality mining dataset at different time slots to find the maximum cliques. The results are shown below. Figure 6 shows sample results with different cliques. The size of the generated clique is based on the time slot of that clique and the number of nodes that interact at this time slot. The results in Figure 7 show the frequency of the discovered maximal cliques at different times for different clique sizes. A total of 12875 cliques were discovered. The 10-clique is the most frequent maximal clique, and the 20-clique is the least frequent as shown in Figure 7.

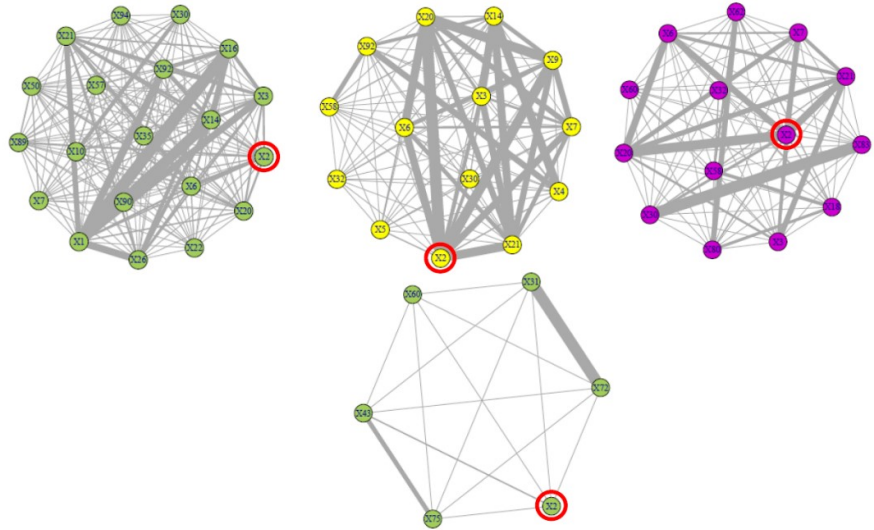

Figure 6. Sample Results of Finding the Maximal Cliques using Bron-Kerbosch Algorithm

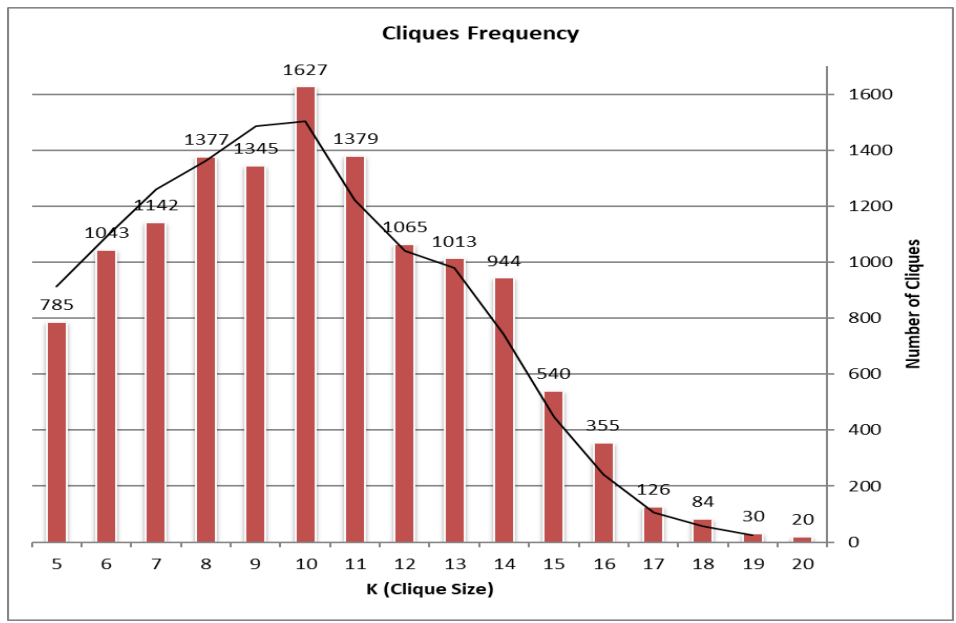

Figure 7. Frequency of Clique Size

The maximal cliques algorithm that was used can be utilized to find the central nodes based on the discovered maximal cliques at every time slot. In this case, the proposed algorithm will be extended to find and generate a list of the common nodes in the discovered maximal cliques at every time step. These nodes are highly connected to most of the other nodes in this case, and they can be considered as central nodes. A time window can be applied in this case as well to find the nodes that frequently continue to be central nodes during this time window. These are the nodes that can be used effectively for early detection of outbreak. For example, the four discovered maximal cliques in Figure 6 at the same time slot can be used to determine the central node. In this example, X2 (with red circle) can be considered as a central node, because it is connected to all other points in the four cliques. X2 can be used in this case as a sensor node to detect an outbreak.

## 7. CONCLUSION

This paper focused on the problem of centrality and maximal cliques in mobile social networks. The goal of this research is to develop evolutionary centrality algorithms that can be applied in mobile social networks to analyze the dynamics of the networks and find the central nodes based on the behavior of the network over time. We also studied the applicability of one of the maximal cliques algorithms. In this paper, we focused on different centrality measures that can be used in mobile social networks with a focus on evolutionary centrality. Finding centrality and maximal cliques in social networks has many important applications including healthcare applications. The proposed evolutionary approach to find centrality in mobile social networks shows a significant enhancement in finding the central nodes using the evolutionary Eigenvector Centrality and PageRank Centrality. The discovered maximal cliques show the dynamic of the different nodes at different time slots, and these maximal cliques can be used as well to find centrality.

Future work includes enhancing the current proposed evolutionary centrality technique by considering more than one time period from previous time slots. Although we were able to discover most of the cliques, there is still a possibility to enhance the results by considering cliques from previous time slots which can be considered for future work.